# Few-cycle mid-infrared pulses from BaGa$_2$GeSe$_6$

Ugaitz Elu[1], Luke Maidment[1*], Lenard Vamos[1], Tobias Steinle[1], Florian Haberstroh[1], Valentin Petrov[2], Valeriy Badikov[3], Dmitrii Badikov[3], and Jens Biegert[1,4]

[1]ICFO - Institut de Ciencies Fotoniques, The Barcelona Institute of Science and Technology, 08860 Castelldefels, Barcelona, Spain.
[2]Max-Born-Institute for Nonlinear Optics and Ultrafast Spectroscopy, 2A Max-Born-Str., D-12489 Berlin, Germany.
[3]High Technologies Laboratory, Kuban State University, Stavropolskaya Str. 149, 350040 Krasnodar, Russia
[4]ICREA - Institució Catalana de Recerca i Estudis Avançats, 08010 Barcelona, Spain.
*Corresponding author: luke.maidment@icfo.eu

**BaGa$_2$GeSe$_6$ (BGGSe) is a newly developed nonlinear material which is attractive for ultrabroad frequency mixing and ultrashort pulse generation due to its comparably low dispersion and high damage threshold. A numerical study shows the material's capacity for octave spanning mid-infrared pulse generation up to 18 µm. In a first experiment, we show that a long crystal length of 2.6 mm yields a pulse energy of 21 pJ at 100 MHz and with a spectral bandwidth covering 5.8 to 8.5 µm. The electric field of the carrier-envelope-phase stable pulse is directly measured with electro optical sampling and reveals a pulse duration of 91 fs which corresponds to sub-4 optical cycles, thus confirming some of the prospects of the material for ultrashort pulse generation and mid-IR spectroscopy.**

Compact and high repetition rate sources of ultrabroadband pulses in the mid-infrared offer many prospects for applications in spectroscopy and remote sensing [1,2]. At the same time, they enable new investigations in such different fields as lightwave electronics, nanoplasmonics, and strong field physics [3]. For all of these fields, particularly enticing are carrier-to-envelope phase (CEP) stable sources since they can provide enhanced sensitivity through field-resolved measurements directly in the time domain. Important for the realization of such sources are nonlinear media amongst which the newly developed BaGa$_2$GeSe$_6$ (BGGSe) crystal is particularly attractive due to its high nonlinearity, favorable dispersion and high damage threshold. BGGSe transmits up to 18 µm and is chemically stable showing no surface aging effects [4]. It can be cleaved in arbitrary directions to the crystal axis and can be antireflection (AR) coated. BGGSe possesses simple trigonal (point group 3) symmetry (uniaxial) and recent characterization of the nonlinear susceptibility tensors shows that the effective nonlinear coefficient that can be used is comparable to AGSe [5]. These properties, coupled with favorably low dispersion between 1.5 - 2.0 µm, make BGGSe a promising candidate for producing high energy, few cycle mid-infrared pulses. For instance, difference frequency generation (DFG) with input wavelengths between 1.7 and 2.6 µm resulted in tunable mid-infrared radiation from 5 to 10 µm [6], and optical parametric amplification yielded pulse energies up to 1.05 µJ at 100 kHz, tunable from 3.9 to 12 µm [7].

Here, we demonstrate the generation of intrinsically CEP-stable sub-4-cycle waveforms at 100 MHz repetition rate and 21 pJ pulse energy from DFG in BGGSe, and we directly measure the electric field waveform of the ultrashort mid-infrared pulse with electro-optical sampling (EOS). To investigate the prospects of ultrashort pulse generation and its wavelength range, we calculate the matched signal and idler wavelengths for a pump wavelength of 1.56 µm. Figure 1(a) shows the result, indicating that phase matching exists across a broad range in the mid-infrared by varying θ between 24 and 30 degrees. As BGGSe is a positive uniaxial crystal the polarizations follow the relation: o(1.56) -> e(2.0) + e(7.1). Equally important for ultrashort pulse generation and efficient DFG, Fig. 1(b) shows the pulse splitting length and gain bandwidth for a pump pulse duration of 60 fs. The comparison with GaSe clearly shows the favorable aspects of BGGSe since the medium allows the use of a longer crystal than GaSe before temporal walk off becomes a limiting factor. At the same time, it provides a larger gain bandwidth.

For the experiment, we use a 100-MHz multi-arm, multi-wavelength master oscillator power amplifier (MOPA) system from Menlo Systems. The MOPA system features a common modelocked Er:fiber oscillator at 100 MHz after which the output is split into several arms. One of these arms further amplifies the at 1.56 µm radiation to 0.5 W and compresses to 59 fs. Another arm incorporates a highly nonlinear fiber whose Raman soliton is shifted to 2.0 µm and further amplified in Tm:Ho fiber . The 2.0-µm output is compressed in a grating compressor to 215 fs and then overlapped with the 1.56-µm pulses using a dichroic mirror. The spectral and temporal profiles of the two near-infrared pulses are measured using second harmonic generation frequency resolved optical gating (SHG FROG). Figure 2 shows a sketch of the setup together with the electro optical sampling (EOS) diagnostics. We recombined the two-color output from the MOPA system inside the BGGSe crystal for difference frequency generation of CEP-stable mid-infrared pulses.

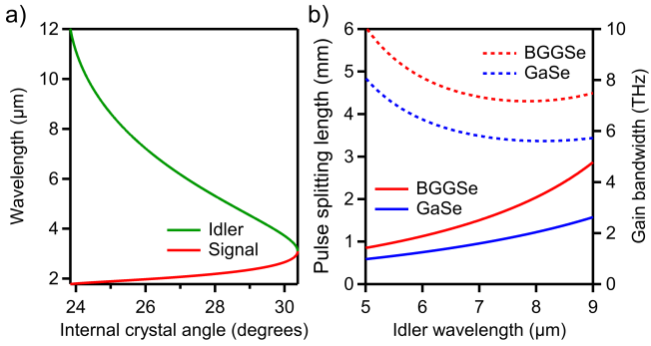

Figure 1: Phase matching curve of the signal and idler wavelengths against internal BGGSe θ angle for 1.56-μm pump. (b) Pulse splitting length for 60 fs pump pulse and gain bandwidth comparison for BGGSe and GaSe. Calculations for BGGSe use the Sellmeier coefficients refined by Kato *et al.* [8].

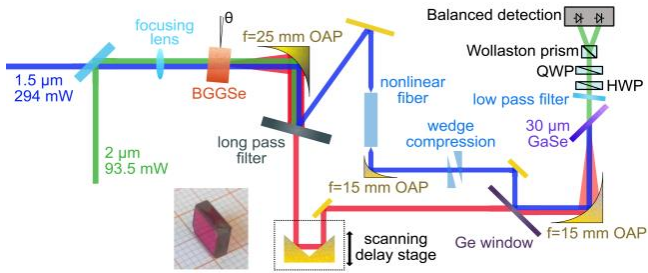

Figure 2: Diagram showing the optical layout of the DFG and EOS measurements, with a photograph of the BGGSe crystal. Deriving both beams for DFG and the EOS sampling pulse from the same oscillator keeps the system compact and optically synchronized.

The BGGSe crystal used has an aperture of 6.48 mm x 5.62 mm, a length of 2.60 mm and is cut at θ = 25°, φ = 30, close to the optimum phase matching angle for type 1 collinear DFG. It has a broadband AR coating for mid-infrared wavelengths. A range of lenses (25, 50 and 75-mm focal length) are used to focus the 1.56- and 2.0-μm beams to different sizes in the BGGSe crystal. In each case the beam size at the focal point was measured using a scanning slit beam profiler. From these measurements the 1.56-μm peak intensity was calculated as 4.2, 2.9 and 0.9 GW·cm$^{-2}$ respectively. The Rayleigh length in BGGSe for 1.56 μm in each case is also determined to be 1.4, 2.1 and 6.2 mm respectively. The mid-infrared difference frequency (idler) beam is collimated with a 25-mm focal length gold off-axis parabolic (OAP) mirror and a dichroic beam splitter is used to separate it from the near infrared seed beams.

The maximum DFG efficiency is achieved using the 50-mm lens producing a pulse energy of 21 pJ (2.1-mW average power). Figure 3(a) shows the spectrum in blue, spanning 2400 nm at a -20 dB level, which has a transform limited duration of 67 fs (3 optical cycles). We also have measured the spectrum generated in a 1-mm long uncoated gallium selenide crystal with the same input beam parameters (all polarizations are reversed as GaSe is a negative uniaxial crystal) shown in red. Similar careful optimization was carried out to maximize the output power and spectral bandwidth, though the angle is optimized for phase matching (12.6° internal angle). The bandwidth of the BGGSe spectrum is wider, despite using a longer crystal. Additionally, only 5 pJ could be generated with GaSe, compared to 21 pJ with BGGSe. The difference is due to the Fresnel reflection losses on both surfaces, as well as the longer interaction length in BGGSe. Despite its greater length, the crystal angle tolerance for the 2.6 mm BGGSe is 11.4 mrad, larger than the tolerance of 9.3 mrad for GaSe.

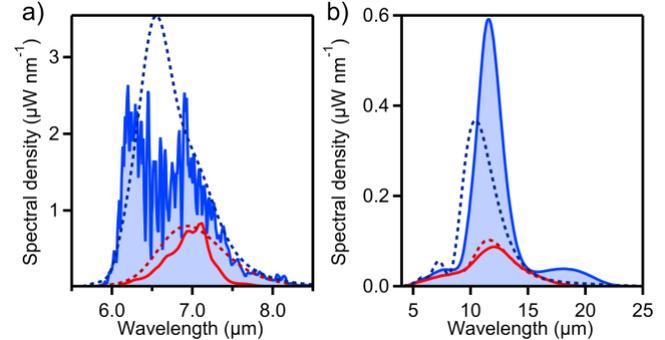

Figure 3: (a) Measured mid-infrared spectrum generated in BGGSe (shadowed blue) and in GaSe (red). Simulated spectra are shown for BGGSe (dotted blue) and GaSe (dotted red). (c) Simulated broadband DFG in 1 mm BGGSe (dotted blue), 2.6 mm BGGSe (shadow blue), 1 mm GaSe (dotted red) and 2.6 mm GaSe (red).

In addition to the experimental results, we simulate the processes with the SISYFOS code [9] based on experimental input parameters such as spectra, energies, and pulse durations. Figure 3(a) shows very good match with the experiment. To further investigate the potential application of BGGSe at wavelengths beyond 12 μm, we also simulated mixing of the 1.56 μm, 59 fs pulses with spectrally broad 30 fs pulses with 2.0 μm center wavelength. The results at the angle of maximum conversion efficiency at 12 μm for 2.6 mm BGGSe and 1.0 mm GaSe are shown in Fig. 3(b). As expected for BGGSe, despite a longer interaction length of 2.6 mm, the output spectrum is as wide as for a 1 mm GeSe, but BGGSe provides higher gain and thus higher conversion efficiency.

We now turn to the temporal characterization of the output from BGGSe and Fig. 2 shows the EOS measurement setup. After transmission through the AR coated BGGSe crystal, the 1.56-μm was coupled into a 22-cm-long, 2.1-μm core highly nonlinear all normal dispersion photonic crystal fiber to generate a 19.0-mW output supercontinuum (not possible with the uncoated GaSe due to the high reflection losses). Fused silica wedges compress the pulses to 20-fs duration (measured with SHG FROG). After the uncoated Ge combiner window, 400 μW (4 pJ) of mid-infrared and 4 mW (40 pJ) of 1.5 μm are spatially overlapped and focused into a 30-μm-thick GaSe crystal angled at 47° (~15.7° internal angle) for type 2 sum-frequency generation (SFG). The mid-infrared beam fills the aperture of the f = 15 mm parabolic mirror to focus as tightly as possible. A nanometer-precise delay stage (SmarAct GmbH) is used for scanning the temporal delay. Interference between the SFG and the sampling pulse results in a polarization rotation for the overlapping spectral components, which is linearly proportional with the mid-infrared electric field [10], and detected by measuring ellipsometry using the combination of waveplates, Wollaston prism and balanced InGaAs detector (Thorlabs PDB210C). We chop the signal at 1 kHz for lock-in detection of the detected signal. Although the 7-μm carrier wavelength with 23-fs cycle duration requires a sample pulse duration below 10 fs, spectral filtering of wavelengths above the generated sum-frequency spectra (1450 nm in this case) increases the temporal resolution by a factor of two, allowing sampling by 20 fs pulses and additionally it improves the SNR by reducing the shot noise [10,11]. To the best of our knowledge this is the lowest energy sample pulse (40 pJ) used for EOS detection, successfully sampling the field of few-pJ mid-infrared pulses.

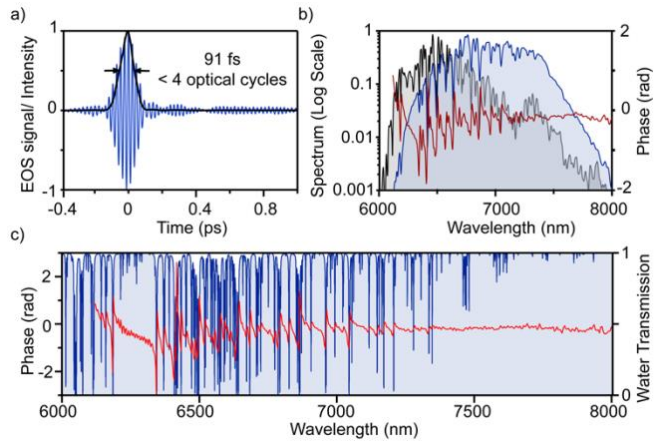

Figure 4: (a) Electric field of the CEP stable mid-infrared pulse generated in BGGSe measured with EOS, averaged over 10 scans. (b) Spectrum from FTIR (grey) compared with EOS spectrum with 2-cm$^{-1}$ resolution (blue) and phase (red). (c) Spectral phase (red) and water transmission (grey) [12].

The fidelity of the measurement is demonstrated by averaging only ten consecutive EOS scans, shown in Fig. 4(a), with the temporal intensity profile overlaid, showing a full width half maximum (FWHM) pulse duration of 91 fs. The main pulse is followed by smaller amplitude oscillations due to the free-induction decay by water vapor (this measurement was carried out at 33% relative humidity). The spectrum calculated from the square of the Fourier transform of the electric field is compared with a simultaneous FTIR measurement in Fig. 4(b). The small difference is due to the remaining phase mismatch, dispersion and absorption, causing only minor changes in the time domain [13]. The spectral phase extracted from the EOS measurement is displayed in Fig. 4(c) and shows excellent agreement with known water absorption lines. Further, the measurement has sufficient resolution and sensitivity to resolve the phase response from individual water absorption lines. Despite the stage using the stick-slip technique (in low vibration mode) and the relatively long beam paths of 2.5 m and 5.5 m, the standard deviation of the zero-crossing fluctuations [14] were below 1.5 fs. The fact that the averaged EOS signal still shows a clear field resolved waveform is testimony to the CEP stability of the sub-4- cycle mid-infrared pulse.

In summary, we have investigated the newly developed nonlinear crystal BGGSe with respect to the generation of ultrashort and ultrabroadband mid-infrared radiation. We used a 2.6 mm long BGGSe crystal for DFG of the two-color output of a fiber MOPA system at 100 MHz. Electro optical sampling is used to directly measure the electric field waveform of the generated mid-infrared pulse which is CEP-stable and exhibits a duration of sub-4 optical cycles or 91 fs. Simulations further indicate the material's utility to generate octave spanning spectra and shorter pulse durations. These aspects, combined with the possibility to grow larger aperture crystals, feature BGGSe as a promising new material for high energy and short pulse generation in the mid-infrared spectral region.

**Acknowledgements**

We acknowledge financial support from the European Research Council for ERC Advanced Grant "TRANSFORMER" (788218), ERC Proof of Concept Grant "miniX" (840010), FET-OPEN "PETACom" (829153), FET-OPEN "OPTOlogic" (899794), Laserlab-Europe (654148), Marie Sklodowska-Curie ITN "smart-X" (860553), MINECO for Plan Nacional FIS2017-89536-P; AGAUR for 2017 SGR 1639, MINECO for "Severo Ochoa" (SEV- 2015-0522), Fundació Cellex Barcelona, the CERCA Programme / Generalitat de Catalunya, Army Research Laboratory (W911NF-17-1-0565) and the Alexander von Humboldt Foundation for the Friedrich Wilhelm Bessel Prize.